\documentclass{sigplanconf}

\usepackage{amsmath}
\usepackage{graphicx}

\begin{document}

\setlength{\pdfpageheight}{\paperheight}
\setlength{\pdfpagewidth}{\paperwidth}

\conferenceinfo{PROMOTO '13}{Oct 27, 2013, Indianapolis, IN, USA} 
\copyrightyear{2013} 
\copyrightdata{978-1-nnnn-nnnn-n/13/10} 
\doi{nnnnnnn.nnnnnnn}




\titlebanner{banner above paper title}        
\preprintfooter{short description of paper}   

\title{Touch-enabled Programming for the Lab of Things}

\authorinfo{Zheng Dong}
           {Indiana University}
           {zhdong@indiana.edu}
\authorinfo{Arjmand Samuel}
           {Microsoft Research}
           {arjmands@microsoft.com}

\maketitle

\begin{abstract}
Lab of Things (LoT, lab-of-things.com) is a research platform for interconnection, programming, and large scale deployment of devices and sensors. These devices and sensors can then be used for deployment of field studies in a variety of research areas including elderly care, energy management, and the like. LoT is built on top of HomeOS, a middle-ware component, making interconnection of a wide range of devices possible. LoT also provides cloud storage and remote monitoring capabilities. Traditionally programming on the LoT platform has been done using C\# in Microsoft Visual Studio. While LoT programs developed on the .NET framework offer a rich set of functionality, writing programs on LoT can be challenging for developers who are not experienced with the technology involved. In this demonstration, we introduce an innovative programming approach on the LoT platform by building a Generic Application and creating corresponding libraries on the user-friendly TouchDevelop (touchdevelop.com) programming environment. As an example, we implemented the same functionality of the Lab of Things Alerts application using the new Generic App. In addition to a touch-enabled programming environment, the new approach also significantly saves time and effort developers have to devote when creating a customized Lab of Things application. 
\end{abstract}




\section{Introduction}
Today, a large number of devices and sensors are deployed in homes, and other places where people live, work and play. Sensors can be installed on doors, windows, water heaters, and more; there are sensors detecting temperature, humidity, motion and distance. With the ubiquity of sensors and devices, there is a growing need to effectively interconnect and manage them. There is also a need to monitor status and collect data from these devices and sensors. The challenges of deploying devices and sensors at scale for field studies grow exponentially with increase in number of devices and geographic dispersion. 

Lab of things (LoT)~\cite{HomeLab} (lab-of-things.com) offers a powerful framework to deploy field studies across geographies and enables easy data collection for storage in the cloud. In addition, the framework offers users the capability of remote monitoring system status, and updating installed applications remotely. As a fundamental local component of LoT, HomeOS middle-ware runs on a local machine~\cite{HomeOS, HomeOSNSDI}, which is usually a small laptop computer, known as the HomeOS hub. The HomeOS hub manages the interconnection with devices and sensors through device drivers, and allows LoT applications to access functionality through a set of APIs.  

The Lab of Things applications follow the Browser/Server (B/S) architecture. The core functionality of LoT applications is written using the WCF~\cite{WCF} web services. Traditionally, these web services have been written in C\# with Microsoft Visual Studio. Following the standard method of creating the WCF service, an app developer must complete two C\# files: \textbf{IService.cs} and \textbf{Service.cs}. The first file defines the web service, the functionality of the application as \textit{endpoints} and a number of web function calls as \textit{operation contracts}. The second file implements the function calls as defined in the first file, and is required to include functions to respond to five basic HomeOS events: \textit{when the platform starts}, \textit{when the platform stops}, \textit{when a new device is registered with the platform}, \textit{when an existing device is disconnected from the platform}, and \textit{when a device-specific event is received from the platform}. To enable better user experience, most LoT apps also include an HTML page with the underlying JavaScripts to interact with the web service. The client-side web page provides easy access to application functionality for developers. In summary, developing LoT applications require 1) Expertise of C\# and WCF. 2) Installation of Visual Studio (only available on Microsoft Windows). 3) Understanding of the Lab of Things architecture. 4) Expertise of HTML, JavaScript, JQuery and JSON.

The above set of requirements puts a high bar for a developer to get started with developing LoT applications. It is expected that an entry-level developer will spend several weeks in order to get familiar with the various platforms and technologies involved. Subsequently, programming LoT applications involves writing a few hundred lines of code in C\#, HTML and JavaScript, depending on the features being implemented. The steep learning curve and substantial time investment is required for creating the Lab of Things applications, especially for computation-intense operations such as motion detection or complex tasks that reference third-party libraries.

One example scenario for the Lab of Things is that of a health informatics researcher who needs to obtain the current temperature and humidity in houses of senior citizens. After the data is obtained, she needs these results to be visualized using bar charts for the average temperature and humidity for each month. The logic of this application is quite simple: it involves collecting data from sensors and basic statistical analysis. However, developing such a simple LoT application using the above described .NET framework may require expertise beyond a typical health informatics researcher. There is a need to enable a quick and easy way to develop simple data collection and visualization applications for the Lab of Things. 

\section{Lightweight Programming Layer on LoT}
We propose that this problem can be solved by introducing an innovative programming layer on top of the existing architecture of Lab of Things. This new Lab of Things layer is designed for quickly creating lightweight applications that no longer require the proficiency of C\# and WCF. It also eliminates the need to install Microsoft Visual Studio as programming environment. This approach only leaves one component for the developers to write: the client-side scripts for web service interaction.

To reduce changes to be made on the existing Lab of Thing framework (including the local HomeOS platform), we built a very generic Lab of Things application using .NET technology that exposes a number of frequently used functionality as endpoints. This application is referred to as the Generic App since it is not designed for a specific device or sensor. Instead, it can potentially interact with all devices that are currently connected to the LoT framework. Through a combination of several web function calls, it is as simple as a few lines of code for most simple tasks such as data collection and statistics.

\begin{figure}
\centering
\includegraphics[width=3in]{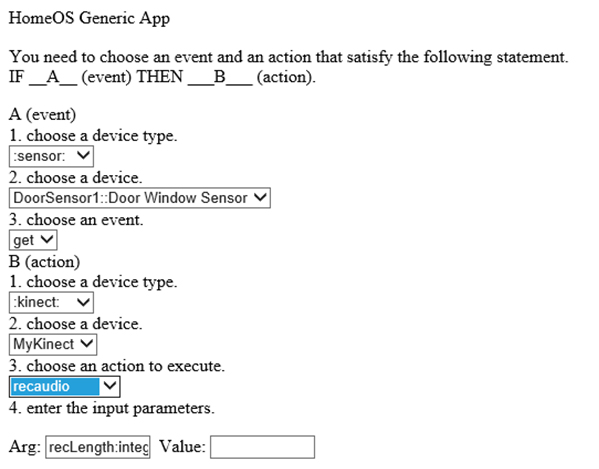}
\caption{Generic App Web Interface for the Lab of Things}
\label{fig:GenericWeb}
\end{figure}

Similar to traditional applications, the Generic App also has a server-side component running on Lab of Things as an individual web service. As mentioned above, developers need to then focus on methods designed for interacting with the generic application. We can envision two ways that such communication can take place. The first method is to access the functionality from the web interface of the Generic App. As shown in Figure~\ref{fig:GenericWeb}, the Generic App web page offers a straightforward process for creating any application with an If-Then logic. Through a number of drop down lists, developers only need to select the devices (\textbf{A} or \textbf{B}), an event (\textbf{Evt}) and an action (\textbf{Act}). When the process finishes, it generates an application that executes \textbf{B.Act} when \textbf{A.Evt} occurs.

The Generic App enables development of lightweight Lab of Things applications. The current web interface of the Generic App, however, only showcases one possible way of utilizing this app. As many developers would imagine, the web interface itself is not flexible enough to support non-If-Then logic. It has not been implemented to display non-textual content (e.g. an image) on the web interface. To enjoy the full features offered by the Generic App, developers can utilize a touch-based programming environment --- TouchDevelop~\cite{TouchDevelop}. TouchDevelop is a web-based programming platform developed by Microsoft Research. It provides a simple app creation environment for mobile, cloud connected and touch devices. Anyone connected to the Internet can visit the \textit{touchdevelop.com} and develop their own scripts. Programming with the Generic App on TouchDevelop can significantly reduce the time and code required for customized Lab of Things applications.

TouchDevelop is among the top candidates for creating client-side scripts for the following reasons.

First, the experience of writing programs has been improved by features such as \textit{auto-completion} and \textit{auto-correction}. Developers do not need to type the entire line of code string, instead only the first few letters are sufficient for TouchDevelop to generate the corresponding candidate commands. Whenever programming errors are suspected, the environment instantly displays suggestions to correct the code. Debugging mode is supported on TouchDevelop and any variable on the stack can be monitored.

There are a number of predefined functions on TouchDevelop for handling multimedia contents (e.g. images, audio clips). Basic functionality such as displaying an image has been made standard functions on TouchDevelop. The programming environment has been optimized for touch-enabled mobile devices such as smartphones and tablets. Developers can write their Lab of Things applications even when they are connected remotely to their HomeOS hubs.

Additionally, TouchDevelop offers built-in capability to make REST-ful~\cite{RESTFul} function calls (e.g. POST and GET). It understands the JSON structure and can even generates one with only a few lines of code. Scripts written on TouchDevelop can be compiled and published for future references. Therefore, with a standard Lab of Things library created on TouchDevelop, one can further improve the programming experience on Lab of Things by linking TouchDevelop (client-side) to the Generic App (server-side). We discuss this in the following preliminary case study.

\section{Preliminary Case Study: Lab of Things \textit{Alerts} App}
Let's consider an elderly care scenario to be implemented using the Lab of Things. As discussed in~\cite{PortalMonitor}, one of the challenges for senior citizens who live alone is social exploitation. While seniors are known to be a vulnerable population to scams, these challenges can be addressed by the Lab of Things Alerts App, a Z-Wave door sensor and a web camera. Technically, whenever the door opens, the LoT Alerts App triggers the web camera that faces the door to take a picture. The image can then be sent to a dedicated caregiver through e-mail. As an optional component, this alert (text and image) can then be uploaded to a cloud storage service. The Alerts App therefore allows a remote caregiver to be aware of any suspicious visitors to the elderly's house.

\begin{figure}
\centering
\includegraphics[width=3in]{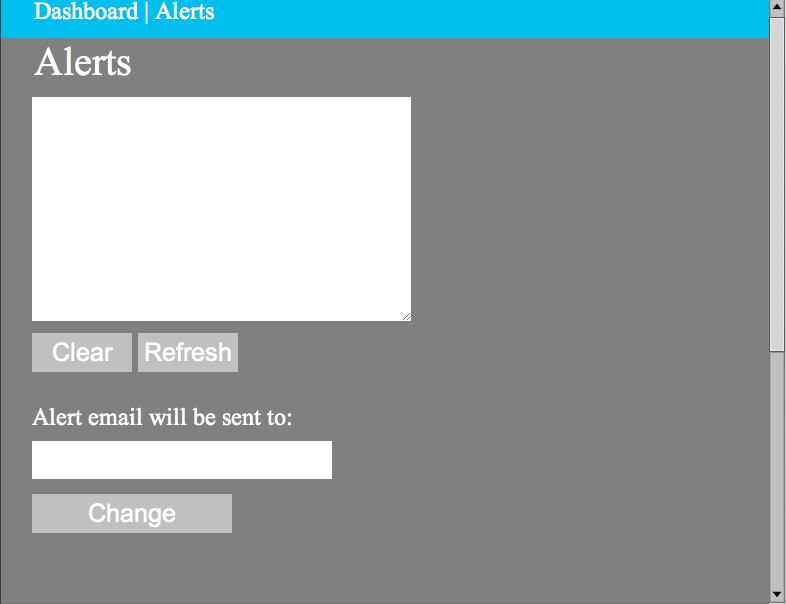}
\caption{Native Alerts App}
\label{fig:NativeAlerts}
\end{figure}

The native Alerts App was written in C\#, HTML and JavaScript. The C\# code has 250 lines in \textbf{AppAlertSvc.cs} (definition of the web service) and 650 lines in \textbf{AppAlerts.cs} (Implementation of the web service). Apart from the standard JQuery and LoT libraries, developers also wrote 220 lines in JavaScript and 125 lines in HTML for a simple web interface as shown in Figure~\ref{fig:NativeAlerts}.

\begin{figure}
\centering
\includegraphics[width=3in]{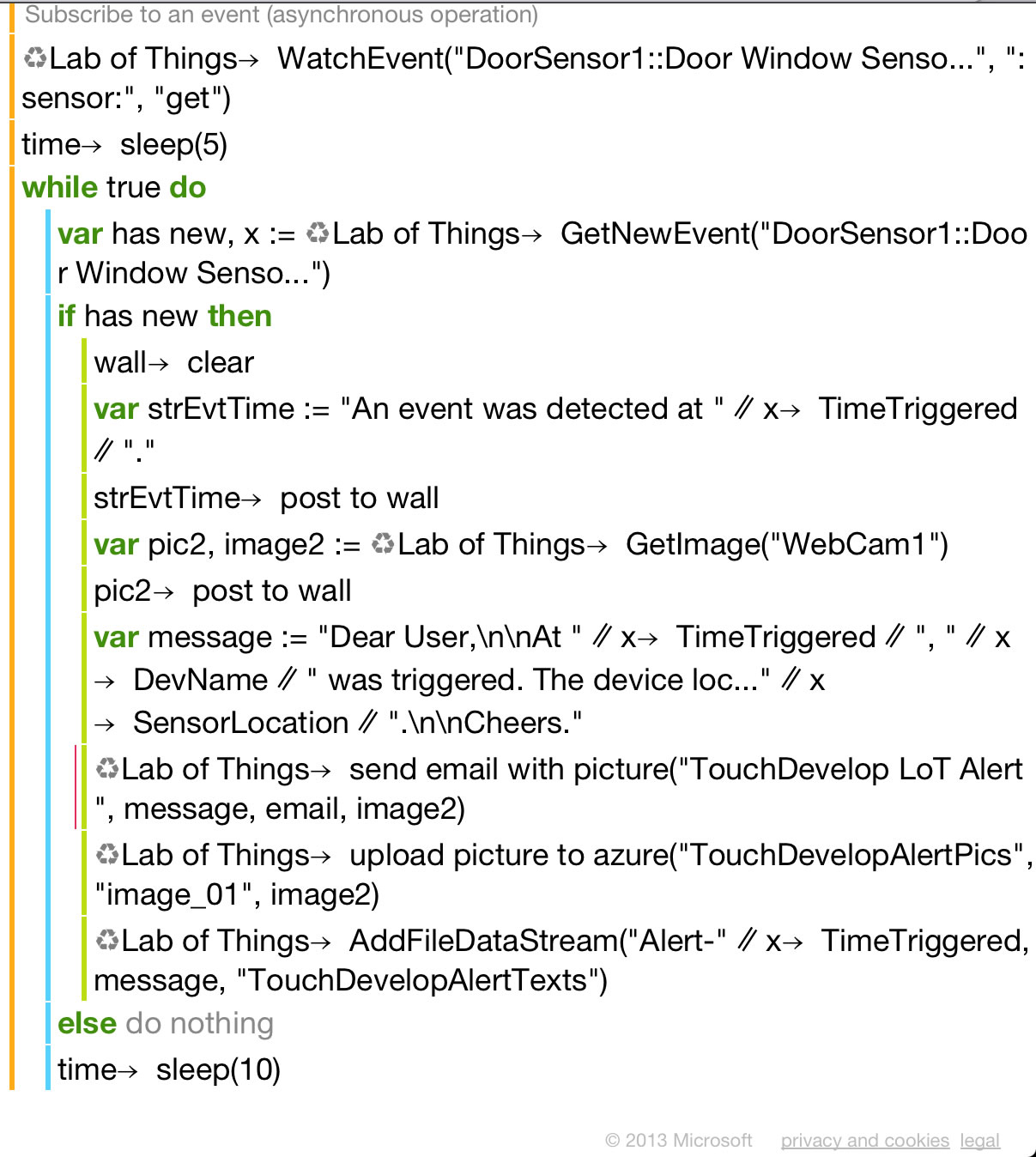}
\caption{TouchDevelop Alerts App}
\label{fig:TDAlerts}
\end{figure}

Fortunately, by utilizing the LoT library on TouchDevelop, we can achieve the same functionality with fewer lines of code. By using this method, a developer can finish the Alerts app with only 15 lines of code on TouchDevelop, as shown in Figure~\ref{fig:TDAlerts}. The TouchDevelop script communicates with the Generic App web service on the lightweight programming layer running on the local HomeOS hub. TouchDevelop Alerts App needs to make the following function calls: \textbf{WatchEvent} for subscribing to future events, \textbf{GetNewEvent} for receiving the latest status updates, \textbf{GetImage} for retrieving photos from camera, \textbf{SendEmailWithImage} for sending emails with an image as the attachment, \textbf{UploadPictureToAzure} for uploading the photo to a cloud storage and \textbf{AddFileDataStream} for updating the alert texts to the remote service.

This is only a preliminary case study. We expect to further our work and publish a full study in due course of time.

\section{Conclusion}
In this work, we introduce an approach to creating applications for the Lab of Things (LoT) platform. Instead of relying on the .NET Framework and requiring expertise of C\# and the installation of Microsoft Visual Studio. We introduce a lightweight programming layer on top of all devices, sensors and traditional applications. With a generic web service running on HomeOS hub, and the LoT library on TouchDevelop, the size of a LoT application can be greatly reduced to only a few lines of code. With a preliminary case study of LoT Alerts App, we demonstrated that Lab of Things applications whereby simple application logic can be effectively developed on touch-based mobile devices using TouchDevelop. 






\bibliographystyle{abbrvnat}




\bibliography{ProMoToBib}

\end{document}